\documentclass[acus]{JAC2003}

\usepackage{graphicx}
\def\ppg{\pi^{+}\pi^{-}\gamma}
\def\ppp{\pi^{+}\pi^{-}\pi^{0}}
\def\eeg{e^{+}e^{-}\gamma}
\def\mmg{\mu^{+}\mu^{-}\gamma}
\def\ifm#1{\relax\ifmmode#1\else$#1$\fi}
\def\epm{\ifm{e^+e^-}} 
\def\to{\ifm{\rightarrow}}
\def\dif{\hbox{d}} \def\sig{\ifm{\sigma}} 
\def\gam{\ifm{\gamma}}  \def\x{\ifm{\times}}
\def\sprime{\ifm{s}'}

\def\up#1{$^{#1}$}  

\def\B{Bari}
\def\b{\rlap{\kern.2ex\up a}}

\def\o{\rlap{\kern.2ex\up b}}

\def\fr{\rlap{\kern.2ex\up c}}
\def\Ka{Karlsruhe}
\def\ka{\rlap{\kern.2ex\up d}}
\def\Le{Lecce}
\def\le{\rlap{\kern.2ex\up e}}

\def\mo{\rlap{\kern.2ex\up f}}
\def\N{Napoli}
\def\n{\rlap{\kern.2ex\up g}}
\def\co{\rlap{\kern.2ex\up i}}
\def\novo{\rlap{\kern.2ex\up h}}
\def\Pi{Pisa}
\def\pI{\rlap{\kern.2ex\up i}}

\def\ra{\rlap{\kern.2ex\up k}}
\def\en{\rlap{\kern.2ex\up j}}

\def\rb{\rlap{\kern.2ex\up l}$\,$}

\def\rc{\rlap{\kern.2ex\up m}}
\def\su{\rlap{\kern.2ex\up n}}
\def\T{Trieste}
\def\t{\rlap{\kern.2ex\up o}}

\def\vi{\rlap{\kern.2ex\up p}}

\def\aff#1{Dipartimento di Fisica dell'Universit\`a e Sezione INFN, #1, Italy.}
\def\hsa{ \ }
\def\hsb{\hskip 2.8mm}

\begin{document}

\title{MEASUREMENT OF $\sigma(e^{+}e^{-} \rightarrow \pi^+\pi^-)$ AT DA$\Phi$NE WITH THE RADIATIVE RETURN}

\author{\baselineskip 13pt
A.~Aloisio\n,\hsa
F.~Ambrosino\n,\hsa
A.~Antonelli\fr,\hsa 
M.~Antonelli\fr,\hsa 
C.~Bacci\rc,\hsa
G.~Bencivenni\fr,\hsa
S.~Bertolucci\fr,\hsa \and
\leftline{C.~Bini\ra,\hsa
C.~Bloise\fr,\hsa 
V.~Bocci\ra,\hsa
% ...
}
\\
\\
 \baselineskip=12pt
\parskip=0pt
\parindent=0pt
\leftline{\b\hsb \aff{\B}} \\
\leftline{\rlap {\kern.2ex\up b} \hsb Permanent address: Institute of High Energy Physics of Academica Sinica, 
Beijing, China.} \\
\leftline{\fr\hsb  Laboratori Nazionali di Frascati dell'INFN, Frascati, Italy.} \\
\leftline{\ka\hsb  Institut f\"ur Experimentelle Kernphysik, Universit\"at \Ka, 
Germany.} \\
\leftline{\le\hsb \aff{\Le}} \\
\leftline{\mo\hsb Permanent address: Institute for Theoretical and Experimental Physics, Moscow,
Russia.} \\
\leftline{\n\hsb Dipartimento di Scienze Fisiche dell'Universit\`a ``Federico II'' e 
Sezione INFN, \N, Italy.} \\
\leftline{\novo\hsb Permanent address: Budker Institute of Nuclear Physics, Novosibirsk, Russia} \\
\leftline{\pI\hsb \aff{\Pi}} \\
\leftline{\en\hsb Dipartimento di Energetica dell'Universit\`a ``La Sapienza'', Roma, Italy.} \\
\leftline{\ra\hsb Dipartimento di Fisica dell'Universit\`a ``La Sapienza'' e Sezione INFN,
Roma, Italy} \\
\leftline{\rb\hsb Dipartimento di Fisica dell'Universit\`a ``Tor Vergata'' e Sezione INFN,
Roma, Italy} \\
\leftline{\rc\hsb Dipartimento di Fisica dell'Universit\`a ``Roma Tre'' e Sezione INFN,
Roma, Italy} \\
\leftline{\su\hsb Physics Department, State University of New York 
at Stony Brook, USA.} \\
\leftline{\t\hsb \aff{\T}} \\
\leftline{\vi\hsb Physics Department, University of Virginia, USA.} \\
} 

\maketitle

\begin{abstract}
The measurement of hadronic cross sections at low centre-of-mass energies is currently attracting much interest because of the  r\^o{}le these measurements play in the  theoretical evaluation of the anomalous magnetic moment of the muon.  The KLOE experiment at the DA$\Phi$NE $\phi$-factory in Frascati aims to determine the cross section of the dominant contribution $e^{+}e^{-} \rightarrow \pi^{+}\pi^{-}$ at energies below 1 GeV exploiting the radiative return from the $\phi$ to the $\rho$ and $\omega$ mesons. This method has systematic errors completely different from previous measurements and allows a cross check of existing data in this energy region. The method is presented in detail and results on both the differential cross section for  $e^{+}e^{-} \rightarrow \pi^{+}\pi^{-}\gamma$ and the total cross section for  $e^{+}e^{-} \rightarrow \pi^{+}\pi^{-}$ are presented.
\end{abstract}

\section{INTRODUCTION}

\subsection{Motivation}
The hadronic contribution to the muon anomaly \mbox{$a_\mu = (g_\mu-2)/2$} can be related to the hadronic cross sections via a dispersion relation 
\begin{equation}
a_\mu({\rm hadr})={1\over4\pi^3}\int_{4 m_{\pi}^2}^{\infty}%
\sigma_{\epm\to{\rm hadr}}(s)K(s)\dif s,
\label{eq:amu}
\end{equation}
where the integral is carried out over the invariant mass squared \textit{s} of the hadronic system. The kernel $K(s)$ is a monotonously varying function that behaves approximately like $1/s$. The annihilation cross section also has an intrinsic $1/s$ behaviour and is largely enhanced around the mass of the $\rho$ meson.  Data at low energies contribute therefore strongly to $a_\mu({\rm hadr})$. Since perturbative QCD fails at energies below $\sim 2.5$ GeV, one has to use experimentally measured cross sections in the dispersion integral for low energy regions.  \\
The most recent experimental value for $a_\mu$ is ~\cite{Bennett:2002} \\
$a_\mu^{exp} =  (11659203 \pm 8)$x$10^{-10}$ [E821], \\
while a recent theoretical evaluation gives ~\cite{Davier:2002} ~\cite{Davier:2003}\\
$a_\mu^{theo} = (11659195.6 \pm 6.8)$x$10^{-10}$ [$\tau$ data]\\
$a_\mu^{theo} = (11659180.9 \pm 8.0)$x$10^{-10}$ [$e^+ e^-$ data]\\
The first theoretical evaluation of $a_\mu$ is obtained with $\tau$ decay data as experimental input to the dispersion integral, using conservation of vector current and isospin symmetry to relate $\tau$ decays to \epm annihilation cross sections. The second theoretical evaluation uses \epm data exclusively, including the reanalyzed data from Novosibirsk (\mbox{CMD-2},~\cite{Akhmetshin:2003}). Comparing the theoretical evaluations with the experimental value, one finds a deviation of 2.0 $\sigma$ for the \epm-based theory value, while the $\tau$-based value shows only a 0.7 $\sigma$ effect. In order to resolve this ambiguous situation and to be able to claim an eventual violation of the standard model, more and better information on hadronic cross sections is needed. For a more detailed discussion, see also ~\cite{DeRafael:2003}.
\subsection{Radiative Return}
The standard way to measure hadronic cross sections in the past was to perform an energy scan, in which the energy of the colliding beams was changed to the desired value. The Frascati \epm collider DA$\Phi$NE, however, was designed as a particle factory operating at the fixed energy of the $\phi$ resonance (1020 MeV) with high luminosity and the current physics program does not allow to vary the beam energy away from the $\phi$ resonance. As a consequence of this, the idea has been worked out to obtain $\sig(\epm\to{\rm hadrons})$ at DA$\Phi$NE using the radiative process  \epm\to hadrons + \gam, where the photon has been radiated in the initial state by one of the electron or the positron, lowering the collision energy ~\cite{Spagnolo:1998}~\cite{Binner:1999}.  By measuring this \textit{radiative return} process the hadronic cross sections become accessible from the $\phi$ mass down to the two-pion threshold.  The total cross section $\sigma_{\epm\to{\rm hadrons}}(s)$ can be obtained by measuring $\dif\sigma_{\epm\to{\rm hadrons}+\gam}/\dif s'$, where $s'=M_{\rm hadr}^2$ from the radiative return. The two quantities can be related by the radiator $H$:
\begin{equation}
{\dif\sigma({\rm hadrons}+\gamma)\over\dif s'}\cdot s' =
\sigma({\rm hadrons},s) \x H(s',\theta_\gamma)
\label{eq:H}
\end{equation}
$H$ depends on \textit{s'} and on the polar angle of the radiated photon $\theta_\gamma$. An accuracy better than 1\% is needed for $H$ in order to perform a precision measurement. Radiative corrections have been computed by different groups up to next-to-leading order for the exclusive final hadronic states $\pi^+ \pi^- \gamma$ ~\cite{Gehrmann:2001} \cite{Hoefer:2001} \cite{Khoze:2002} and  4$\pi+\gamma$ ~\cite{Czyz:2001}.
Our analysis makes use of the theoretical MonteCarlo event generator PHOKHARA ~\cite{Gehrmann:2001}~\cite{Rodrigo:2001}~\cite{Kuhn:2002}~\cite{Rodrigo:2002}, which also includes now final state radiation of the pions using scalar QED ~\cite{Czyz:2003}.\\
Below 1 GeV, the contribution of the process  \mbox{$e^+ e^- \to \pi^+ \pi^- $} to the total hadronic contribution to $a_\mu$ is 67\%~\cite{Jegerl:2003}. The cross section for this process can be obtained using eq.~\ref{eq:H} by measuring $\dif\sigma(\epm\to{\rm \ppg})/\dif s'$. In the following results obtained for this process with the KLOE detector in Frascati will be reported. 

\section{Measurement of $\dif\sigma(\epm\to{\rm \ppg})/\dif \sprime $}
\subsection{Signal selection}
\begin{figure}
\vspace{6pt}
\includegraphics[width=17pc]{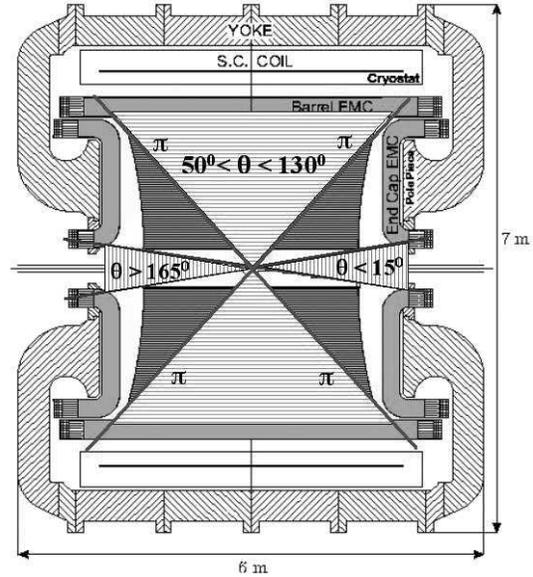}\vglue-.5cm
\caption{Schematic view of the KLOE detector with the angular acceptance regions for pions (\textit{horizontally hatched area}) and photons (\textit{vertically hatched area}). The photon angle is evaluated from the two pion tracks.}
\label{fig:fiducial}
\end{figure}
\begin{figure}[t] 
\vspace{6pt}
\includegraphics[width=17pc]{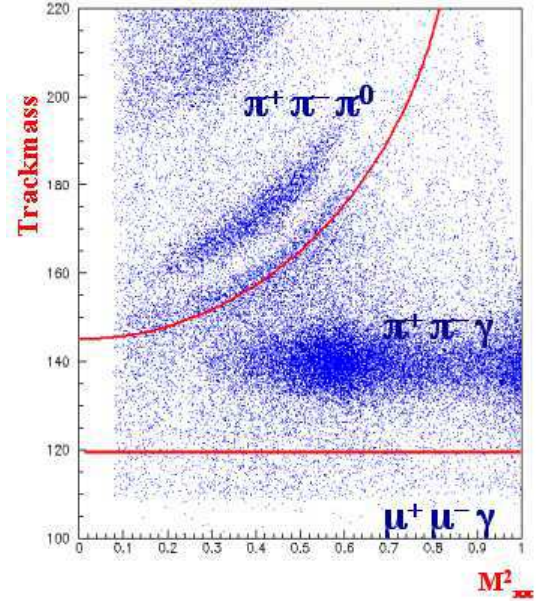}\vglue-.5cm
\caption{2-dimensional cut in the plane of track mass/MeV and $M_{\pi\pi}^2$/GeV$^2$ }
\label{fig:trackmass}
\end{figure}

The KLOE detector consists of a high resolution drift chamber ( $\sigma_{p_T} / p_T \leq 0.4\%$,~\cite{kloedc:2002}) and an electromagnetic calorimeter ($\sigma_E / E = 5.7 \% /  \sqrt{E({\rm GeV})}$,~\cite{kloeemc:2002}). In the current analysis we have concentrated on events in which the pions are emitted at polar angles $\theta_\pi$ between $50^\circ$ and $130^\circ$. No explicit photon detection is required and an \textit{untagged} measurement is performed. As a consequence, a cut on the di-pion production angle $\theta_{\pi\pi}$ smaller than $15^\circ$ (or greater than $165^\circ$) is performed instead of a cut on the photon angle $\theta_\gamma$. $\theta_{\pi\pi}$ is calculated from the momenta of the two pions and is exactly equal to $180^{\circ}-\theta_\gamma$ if only one photon is emitted. The acceptance regions are shown in Fig.~\ref{fig:fiducial}. The reason to choose these specific acceptance cuts are reduced background contaminations and a low relative contribution of final state radiation from the pions. It will be shown in the following that an efficient and almost background free signal selection can be performed without explicit photon tagging ~\cite{Cataldi:1999} \cite{smueller:2003} \cite{incagli:2003}.  \\
The selection of $e^{+}e^{-} \rightarrow \ppg$ is performed with the following steps:
\begin{itemize}
\item[-]\textit {Detection of two charged tracks connected to a vertex}: Two charged tracks with polar angles between $50^\circ$ and $130^\circ$ connected to a vertex in the fiducial volume $R<8$ cm, $|z|< 15$ cm are required. Additional cuts on transverse momentum $p_{T} > 200$ MeV or on longitudinal momentum $|p_{z}| > 90$ MeV reject spiralizing tracks and ensure good reconstruction conditions.
\item[-]\textit {Identif{}ication of pion tracks}: A $\pi$-$e$ separation is done by cutting on a likelihood function which parametrizes the interaction of charged particles with the electromagnetic calorimeter. The function has been built using control samples from data of  $\pi^{+}\pi^{-}\pi^{0}$ and $\eeg$ in order to find the calorimeter response for pions and electrons. The event is selected if at least one of the two tracks is identified to be a pion. In this way, $\eeg$ events are drastically removed from the signal sample while the efficiency for $\ppg$ is still very high ($>$ 99\%).
 \item[-]\textit {Cut on the track mass}: The track mass is a kinematic variable corresponding to the mass of the charged tracks  under the hypothesis that the final state consists of two particles with the same mass and one photon. It is calculated from the reconstructed momenta $\vec{p}_{+}$, $\vec{p}_{-}$.  Cutting at a value of 120 MeV rejects $\mmg$ events, while in order to reject  $\pi^{+}\pi^{-}\pi^{0}$ events a cut in the plane of track mass and $M_{\pi\pi}^2$ is performed (see Fig.~\ref{fig:trackmass}):
 \begin{equation}
 m_{track} < 250. - 105.\cdot \sqrt{(1.- \left(\frac{\sprime}{850000.}\right)^2}
\label{eq:trackmass}
 \end{equation}  
 All units in eq.~\ref{eq:trackmass} are in MeV.
 \item[-]\textit {Cut on the di-pion angle $\theta_{\pi\pi}$}: The aforementioned cut on the di-pion angle $\theta_{\pi\pi} < 15^\circ$ or $\theta_{\pi\pi} > 165^\circ$ is performed.  
\end{itemize}
\subsection{Effective cross section}
\begin{figure}[t] 
\vspace{6pt}
\includegraphics[width=17pc]{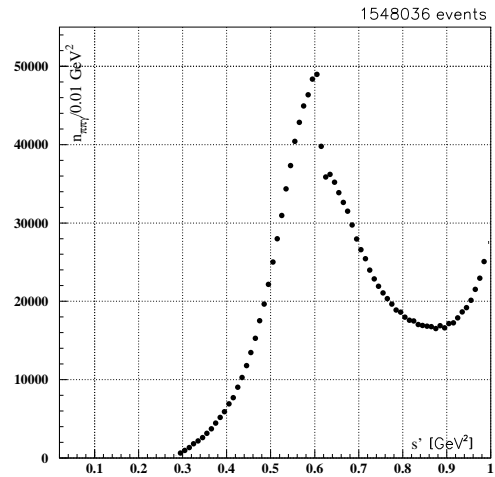}\vglue-.5cm
\caption{Number of $\ppg$ events selected by the analysis in the region 
$\theta_{\pi\pi}<15^\circ$ or $\theta_{\pi\pi}>165^\circ$, $50^\circ<\theta_{\pi}<130^\circ$.}
\label{fig:spectrum}
\end{figure}

Applying the selection steps mentioned in the previous section on a sample of $\sim140$ pb$^{-1}$ of data taken in 2001, ca. 1.5 Mio. events are selected, corresponding to  11000 events/pb$^{-1}$. Fig.~\ref{fig:spectrum} shows the number of $\ppg$ events selected in bins of $0.01$ GeV$^2$. The $\rho-\omega$ interference is clearly visible, demonstrating the excellent momentum resolution of the KLOE drift chamber.
To obtain the effective $\ppg$ cross section one has to subtract the residual background from this spectrum and divide by the selection efficiency and the integrated luminosity:
\begin{equation}
\frac{d\sigma_{\rm \ppg}}{dM_{\pi\pi}^2}=
\frac{dN_{{\rm Obs}}-dN_{{\rm Bkg}}}{dM_{\pi\pi}^2} \: 
\frac{1}{\epsilon_{\rm Sel}} \:
\frac{1}{\int{\mathcal{L}}dt}
\label{eq:ana}
\end{equation}
\begin{itemize}  
\item[-]{\textit{Residual background $N_{Bkg}$}}: Given the angular acceptance regions for pions and photon as described in fig.~\ref{fig:fiducial}, the amount of residual background is considered to be small. It has been evaluated by fitting the MC distributions for $\ppg$ (\textit{signal}) and $\mmg$, $\eeg$ and $\ppp$ (\textit{background}) to the data. The relative contribution of each background channel is found to be well below 5\%.
\item[-]{\textit{The selection \mbox{eff{}iciency} $\epsilon_{\rm Sel}$}}: The selection efficiency is the product of trigger, reconstruction, filtering, $\pi$-$e$ separation and track mass cut efficiencies. Apart from the last one, which has been evaluated from MonteCarlo, all efficiencies have been evaluated from data, using unbiased control samples matching the kinematics of the actual signal. The total selection efficiency is about  $70\%$.
\item[-]{\textit{Luminosity}}: At KLOE, the luminosity measurement is performed using  Bhabha events at large angles ($55^\circ<\theta_{+,-}<125^\circ$, $\sigma=425$nb). The number of Bhabha candidates is counted and the luminosity is obtained from the effective cross section obtained from MonteCarlo. KLOE uses two independent Bhabha event generators (the Berends/Kleiss generator~\cite{Berends:1983} modified for DA$\Phi$NE~\cite{Drago:1997} and the BABAYAGA generator~\cite{Calame:2000}), for both of which the authors quote a theoretical error of 0.5\%. Both generators agree within $0.2\%$. The good agreement of the experimental distributions ($\theta_{+,-}$, $E_{+,-}$) with the event generators results in an experimental uncertainty for the luminosity measurement of 0.4\%. 
\end{itemize}
\begin{figure}[t] 
\vspace{6pt}
\includegraphics[width=17pc]{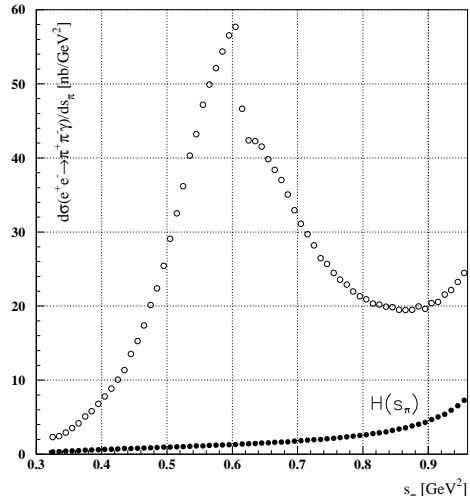}\vglue-.5cm
\caption{Differential cross section for $e^+e^-\rightarrow \pi^+\pi^-\gamma$. The radiator $H(\sprime)$ obtained from MC is also shown.}
\label{fig:pipig}
\end{figure}
The effective cross section \mbox{$\dif\sigma(\epm\to{\rm \ppg})/\dif \sprime$} within the acceptance cuts described in fig.~\ref{fig:fiducial} is shown in fig.~\ref{fig:pipig}.
\section{Extraction of $\sigma(\epm\to{\rm \pi^{+}\pi^{-}})$}
Assuming that initial and final state radiation processes are independent (cf. neglecting interference effects) the cross section for $\epm\to{\rm \pi^{+}\pi^{-}}$ can be extracted as a function of $M^2_{\pi\pi}$ from the observed differential $\ppg$ cross section as:
\begin{equation}
\sigma(\epm\to{\rm \pi^{+}\pi^{-}})= \frac{\pi \cdot \alpha^2}{3\cdot s} \cdot \beta_{\pi}^3 \cdot \frac{\sigma_{\pi\pi\gamma}}
{\sigma_{\pi\pi\gamma}({F_\pi=1})}
\label{eq:pion}
\end{equation}
where $\sigma_{\pi\pi\gamma}({F_\pi=1})$ is the NLO cross section for $e^{+}e^{-} \rightarrow \ppg$ (initial state radiation only) under the assumption of pointlike pions and corresponds to the quantity $H$ from eq.~\ref{eq:H}. We obtain $H$ bin-by-bin technically from the theoretical MonteCarlo generator PHOKHARA by setting $F_\pi=1$ and switching off the vacuum polarisation of the intermediate photon (see fig.~\ref{fig:pipig}).  $\beta_{\pi}$ is the pion velocity defined as $(1.-4.\cdot m_{\pi}^2 /s)^{\frac{1}{2}}$. Note that no explicit acceptance correction has been performed, but $H$ is evaluated within the acceptance cuts of fig.~\ref{fig:fiducial}. Dividing by the \textit{effective} $H$ thus yields the total cross section for $e^{+}e^{-} \rightarrow \pi\pi$.  
\subsection{Radiative corrections}
\begin{figure}[t] 
\vspace{6pt}
\includegraphics[width=17pc]{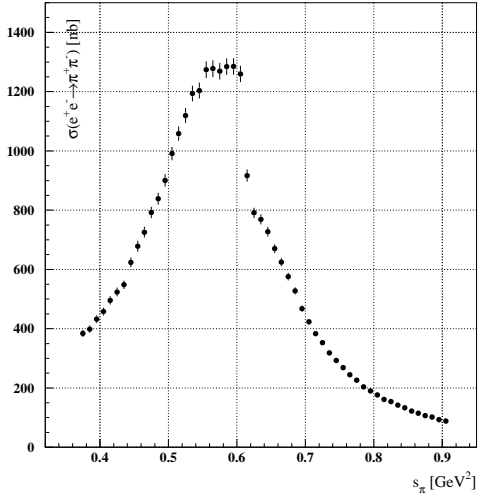}\vglue-.5cm
\caption{Cross section for $e^+e^-\rightarrow \pi^+\pi^-$.}
\label{fig:pipi}
\end{figure}

In order to evaluate $a_\mu({\rm hadr})$ by inserting the cross section in the dispersion integral (eq.~\ref{eq:amu}), some radiative corrections to the cross section have to be considered.
\begin{itemize}  
\item[-]{\textit{Final state radiation}}:
Events with final state radiation from the pions should be included in the cross section to be put in the dispersion integral~\cite{Hoefer:2002}. The leading order final state radiation where there is no initial state radiation (hence the $e^+$ and the $e^-$ collide at the energy of the $\phi$ meson) creates a large tail, which is unwanted in our case and is suppressed very efficiently to the level of few permille using the acceptance cuts described in fig.~\ref{fig:fiducial} via the separation of pion and photon fiducial volumes (the photon emission by the pions is peaked in the pion direction). 
Considering the next-to-leading order process of simultaneous occurence of a photon radiated in the initial state and a photon radiated in the final state, events of this type are rejected by the aforementioned cut in trackmass. The presence of two photons in the event moves the $\ppg(\gamma)$ event towards higher values of trackmass in fig.~\ref{fig:trackmass}, therefore especially at low $M_{\pi\pi}^2$ they are moved into the region which is cut out by the trackmass cut. As of now, we recover these events by evaluating the trackmass efficiency from the Phokhara 3.0 MonteCarlo~\cite{Czyz:2003} which includes the simulation of this kind of events within the model of scalar QED. Dividing the raw spectrum by this efficiency according to eq.~\ref{eq:ana} retains the $\ppg(\gamma)$ events rejected by the trackmass cut.\footnote{This approach is more specific than the one presented at the workshop, which didn't contain any correction for final state radiation. It was made possible by a deeper understanding of final state radiation events resulting from working with the upgrade of the Phokhara generator.} 
\item[-]{\textit{Vacuum polarisation}}:
The cross section has to be undressed from vacuum polarisation~\cite{Hoefer:2002}. This is done by dividing for the running of the fine structure constant:
\begin{equation}
\sigma_{undressed} = \sigma_{dressed} \cdot \frac{\alpha_0^2}{\alpha(s)^2}
\label{eq:vacpol}
\end{equation}
The correction function to $\alpha_0$ has been provided by F. Jegerlehner~\cite{Jeger:2003}.
\end{itemize}
The resulting total cross section $\sigma(\epm\to{\rm \pi^{+}\pi^{-}})$ after these corrections is shown in fig.~\ref{fig:pipi}.  
\subsection{Systematic error}
The contribution of the several analysis items to the systematic error of the measurement is shown in table~\ref{tab_ex}.  Given the excellent resolution of the KLOE drift chamber a dedicated unfolding from the detector resolution seems not necessary. Eventual systematic contributions from this omission are currently under study, but are expected to be small.
\begin{table}
\begin{center}
\begin{tabular}{|l|c|}
\hline
Acceptance & 0.3\% \\
\hline
Trigger & 0.2\% \\
\hline
Tracking & 0.3\% \\
\hline
Vertex & 0.7\% \\
\hline
Reconstruction filter & 0.6\% \\
\hline
Pion - electron separation & 0.1\% \\
\hline
Trackmass cut & 0.2\% \\
\hline
Background subtraction & 0.5\% \\
\hline
\hline
\bf{Total exp. systematic error:} & \bf{1.2\%} \\
\hline
\end{tabular}
\caption{List of exp. systematic uncertainties}
\label{tab_ex}
\end{center}
\end{table}
In table~\ref{tab_theo} the systematic uncertainties coming from theoretical input to the evaluation of $\sigma(\epm\to{\rm \pi^{+}\pi^{-}})$ are listed (for the luminosity, theoretical and experimental error has been added).
\begin{table}
\begin{center}
\begin{tabular}{|l|c|}
\hline
Radiator $H$ & 0.5\% \\
\hline
Vacuum Pol. & 0.1\% \\
\hline
Luminosity & 0.6\% \\
\hline
\hline
\bf{Total theor. systematic error:} & \bf{0.8\%} \\
\hline
\end{tabular}
\caption{List of theor. systematic uncertainties}
\label{tab_theo}
\end{center}
\end{table}  
Additionally, we attribute a preliminary systematic error of 1.\% due to the treatment of final state radiation in our analysis. This covers the limited validity of the radiator $H$ due to eventual non-factorizing terms between initial and final state radiation, possible limitations of the model used to simulate final state radiation, the error arising due to the fact that in the presence of final state radiation the (measured) $s_\pi$ is slightly different from $s_{\gamma^*}$ (where $\gamma^*$ is the virtual photon produced in the $e^+e^-$-collision) and additional acceptance corrections in case the acceptance changes in the presence of final state radiation events.  
\section{Conclusions}
A method of measuring hadronic cross sections at particle factories has been presented using the radiative return. This method is complementary to the energy scan used to perform this measurement up to now and has completely different systematical errors, making it the ideal tool to check and improve existing measurements. Such a crosscheck has become very important given the disagreement between the evaluations of $a_\mu$ obtained in ~\cite{Davier:2002}. 
The radiative return has been used  at the DA$\Phi$NE $\phi$-factory in Frascati in order to extract the total cross section for $e^+e^-\rightarrow \pi^+\pi^-$, making good use of the recently developed upgrade of the Phokhara event generator by~\cite{Czyz:2003}. 
An interpretation of our cross section in terms of $a_\mu({\rm hadr})$ as well as future plans and prospects of this measurement at present and upcoming $e^+e^-$-colliders in Frascati can be found in A. Denig's contribution to this workshop~\cite{Denig:2003}. In particular, the quoted result for the two-pion contribution to $a_\mu({\rm hadr})$ is (in units of $10^{-10}$): 
\begin{eqnarray}
\lefteqn{a_\mu({\rm hadr})_{\pi\pi} = }
\nonumber \\
& & 378.4 \pm 0.8_{stat} \pm 4.5_{syst} \pm 3.0_{theo} + 3.8_{FSR}
\nonumber
\end{eqnarray}
Our preliminary result agrees within the errors with the result found by CMD2~\cite{Akhmetshin:2003}.  
\section{Acknowledgement}
We thank the DA$\Phi$NE team for their efforts in maintaining low background running
conditions and their collaboration during all data-taking.
We want to thank our technical staff:
G.F.Fortugno for his dedicated work to ensure an efficient operation of
the KLOE Computing Center;
M.Anelli for his continous support to the gas system and the safety of the
detector;
A.Balla, M.Gatta, G.Corradi and G.Papalino for the maintenance of the
electronics;
M.Santoni, G.Paoluzzi and R.Rosellini for the general support to the
detector;
C.Pinto (Bari), C.Pinto (Lecce), C.Piscitelli and A.Rossi for
their help during shutdown periods.
This work was supported in part by DOE grant DE-FG-02-97ER41027;
by EURODAPHNE, contract FMRX-CT98-0169;
by the German Federal Ministry of Education and Research (BMBF) contract 06-KA-957;
by Graduiertenkolleg `H.E. Phys. and Part. Astrophys.' of Deutsche Forschungsgemeinschaft,
Contract No. GK 742;
by INTAS, contracts 96-624, 99-37;
and by TARI, contract HPRI-CT-1999-00088.



\begin{thebibliography}{99} 

\bibitem{Bennett:2002}
G.~W.~Bennett {\it et al.}  [Muon g-2 Collaboration],
Phys.\ Rev.\ Lett.\  {\bf 89} (2002) 101804
[Erratum-ibid.\  {\bf 89} (2002) 129903]

\bibitem{Davier:2002}
M.~Davier, S.~Eidelman, A.~H\"ocker and Z.~Zhang,
Eur. Phys. J. C {\bf 27} (2003), 497

\bibitem{Davier:2003}
M.~Davier, S.~Eidelman, A.~H\"ocker and Z.~Zhang,
submitted to Phys. Lett. B
\mbox{[hep-ph/0308213]}

\bibitem{Akhmetshin:2003}
R.~R.~Akhmetshin {\it et al.}  [CMD-2 Collaboration],
\mbox{[hep-ex/0308008]}

\bibitem{DeRafael:2003}
E.~De Rafael, these proceedings

\bibitem{Spagnolo:1998}
S.~Spagnolo,
Eur.\ Phys.\ J.\ C {\bf 6} (1999) 637

\bibitem{Binner:1999}
S.~Binner, J.~H.~K\"uhn and K.~Melnikov,
Phys.\ Lett.\ B {\bf 459} (1999) 279

\bibitem{Gehrmann:2001}
G.~Rodrigo, A.~Gehrmann-De~Ridder, M.~Guilleaume and J.~H.~K\"uhn,
Eur.\ Phys.\ J.\ C {\bf 22} (2001) 81

\bibitem{Hoefer:2001}
A.~H\"ofer, J.~Gluza and F.~Jegerlehner,
Eur.\ Phys.\ J.\ C {\bf 24} (2002) 51

\bibitem{Khoze:2002}
V.~A.~Khoze, M.~I.~Konchatnij, N.~P.~Merenkov, G.~Pancheri, L.~Trentadue and O.~N.~Shekhovtzova,
Eur.\ Phys.\ J.\ C {\bf 25} (2002) 199
and references there

\bibitem{Rodrigo:2001}
G.~Rodrigo, H.~Czy{\.z}, J.~H.~K{\"u}hn and M.~Szopa,
Eur.\ Phys.\ J.\ C {\bf 24} (2002) 71

\bibitem{Kuhn:2002}
J.~H.~K{\"u}hn and G.~Rodrigo,
Eur.\ Phys.\ J.\ C {\bf 25} (2002) 215

\bibitem{Czyz:2001}
H.~Czy{\.z} and J.~H.~K\"uhn,
Eur.\ Phys.\ J.\ C {\bf 18} (2001) 497

\bibitem{Rodrigo:2002}
G.~Rodrigo, H.~Czy{\.z}, J.~H.~K{\"u}hn and A.~Grzeli\'nska,
Eur.\ Phys.\ J.\ C {\bf 27} (2003) 563

\bibitem{Czyz:2003}
H.~Czy{\.z}, A.~Grzeli\'nska, J.~H.~K\"uhn and G.~Rodrigo
[hep-ph/0308312]

\bibitem{Jegerl:2003}
F.~Jegerlehner, Talk at the ``Workshop on Hadronic Cross Sections at Low Energy'' SIGHAD03, Oct 8-10 2003, Pisa

\bibitem{kloedc:2002}
M.~Adinolfi {\it et al.},
Nucl.\ Instrum.\ Meth.\ A {\bf 488} (2002) 51

\bibitem{kloeemc:2002}
M.~Adinolfi {\it et al.},
Nucl.\ Instrum.\ Meth.\ A {\bf 482} (2002) 364

\bibitem{Cataldi:1999}
G.~Cataldi, A.~Denig, W.~Kluge, S.~M\"uller and G.~Venanzoni,
Published in Frascati Physics Series (2000) 569-578

\bibitem{smueller:2003}
S.~M\"uller for the KLOE collaboration, 
Proceedings of the  PHOTON2003 conference, April 7-11 2003, Frascati,
to be published in Nucl. Phys. B (proc. Suppl.)

\bibitem{incagli:2003}
M.~Incagli for the KLOE collaboration,
Proceedings to the EPS, July 17-23 2003, Aachen

\bibitem{Berends:1983}
F.~A.~Berends and R.~Kleiss,
Nucl.\ Phys.\ B {\bf 228} (1983) 537.

\bibitem{Drago:1997}
E.~Drago and G.~Venanzoni,
INFN-AE-97-48.

\bibitem{Calame:2000}
C.~M.~Carloni Calame, C.~Lunardini, G.~Montagna, O.~Nicrosini and F.~Piccinini,
Nucl.\ Phys.\ B {\bf 584} (2000) 459

\bibitem{Hoefer:2002}
A.~H\"ofer, J.~Gluza, S.~Jadach and F.~Jegerlehner,
Eur. Phys. J, C {\bf 28} (2003) 261

\bibitem{Jeger:2003}
F.~Jegerlehner, private communication

\bibitem{Denig:2003}
A.~ Denig, these proceedings

\end{thebibliography}
\end{document}